\documentclass[aps,prl, preprint,showpacs,superscriptaddress,groupedaddress,amssymb ]{revtex4-1}
\usepackage{graphicx, tabularx, dcolumn, bm, epsfig, amsmath, multirow, array}
\usepackage[compact]{titlesec}
\usepackage{esvect, float}
\usepackage{color}
\usepackage{soul}
\sethlcolor{yellow}
\makeatletter

\newcommand{\Rmnum}[1]{\expandafter\@slowromancap\romannumeral #1@}
\newcommand{\dotr}{\mbox{$\boldsymbol{\cdot}$}}
\makeatother
\begin{document}
\title{Bulk photovoltaic effect enhancement via electrostatic control in layered ferroelectrics}
\author{Fenggong Wang}
\email[]{fenggong@sas.upenn.edu}
\author{Steve M. Young}
\author{Fan Zheng}
\author{Ilya Grinberg}
\author{Andrew M. Rappe}
\email[]{rappe@sas.upenn.edu}
\affiliation{The Makineni Theoretical Laboratories, Department of Chemistry, University
of Pennsylvania, Philadelphia, PA 19104--6323}
\date{\today}
\begin{abstract}
The correlation between the shift current mechanism for the bulk photovoltaic effect (BPVE) and the structural and electronic properties of ferroelectric perovskite oxides is not well understood.
Here, we study and engineer the shift current photovoltaic effect using a visible-light-absorbing  ferroelectric Pb(Ni$_{x}$Ti$_{1-x}$)O$_{3-x}$ solid solution from first principles.
We show that the covalent orbital character dicates the direction, magnitude, and onset energy of shift current in a predictable fashion.
In particular, we find that the shift current response can be enhanced via electrostatic control in layered ferroelectrics, as bound charges face a stronger impetus to screen the electric field in a thicker material, delocalizing electron densities.
This heterogeneous layered structure with alternative photocurrent generating and insulating layers is ideal for BPVE applications.
\end{abstract}
\maketitle

Solar energy is a promising long-term solution for future energy supply challenges because it is  both renewable and environmentally friendly~\cite{Maeda06p295}.
To convert solar energy efficiently, low band gap semiconducting materials are needed that can separate the photo-excited charge carriers well for electricity generation or  catalysis~\cite{Choi09p63, Kudo09p253, Wang12p476}.
While traditional charge separation is accomplished through some externally engineered asymmetry as found for example in a  $p$-$n$ junction, ferroelectrics (FEs) provide an alternative way to separate charge by the internal depolarization field or by the bulk photovoltaic effect (BPVE)~\cite{Glass74p233, Kraut79p1548, Chynoweth56p705}. 
In the BPVE under sustained illumination, the electrons are continuously excited to a quasiparticle coherent state that has an intrinsic momentum, generating a spontaneous direct short-circuit photocurrent.
Furthermore, the BPVE is able to create an open-circuit photovoltage that is above the material's band gap, potentially enabling high power conversion efficiences beyond the Shockley-Queisser limit~\cite{Shockley61p510, Ji10p1763}.
Time dependent perturbation theory analysis clearly shows the roles of broken inversion symmetry and spontaneous charge separation in the BPVE through the ``shift current'' mechanism~\cite{Baltz81p5590, Young12p116601}.

\parskip 0pt
However, the effects of material structure and electronic structure on shift current remain unclear, and this hinders shift current engineering for photovoltaics.
Here,  we elucidate the connections between structure, electronic structure, and shift current using a low band gap solid solution Pb(Ni$_{x}$Ti$_{1-x}$)O$_{3-x}$ (Ni-PTO) based on first-principles calculations~\cite{Bennett08p17409, Gou11p205115}.
We construct the Ni-PTO solid solutions as layered systems, by substituting one Ni atom for Ti in the $1\times1\times N$ supercells of bulk PbTiO$_{3}$ (PTO), and then removing the O atom that is adjacent to the Ni atom at its apical position (Fig. S1 in the Supplementary Material).  These layered materials could potentially be synthesized and remain in the perovskite phase even for high $B$-site substitution levels~\cite{Bennett08p17409}.
We choose this recently designed visible-light absorbing ferroelectric material as a prototypical example for this study because its electronic states near the band gap possess a diverse character that enhances the physical and chemical tunabilities of this material.
We show that the shift current response can be tuned substantially via electrostatic control in these compositionally layered ferroelectrics, potentially enabling future high power conversion efficiency. 

\begin{figure}[b]
\centering
\vspace{-1.0em}
\includegraphics[width=0.8\textwidth]{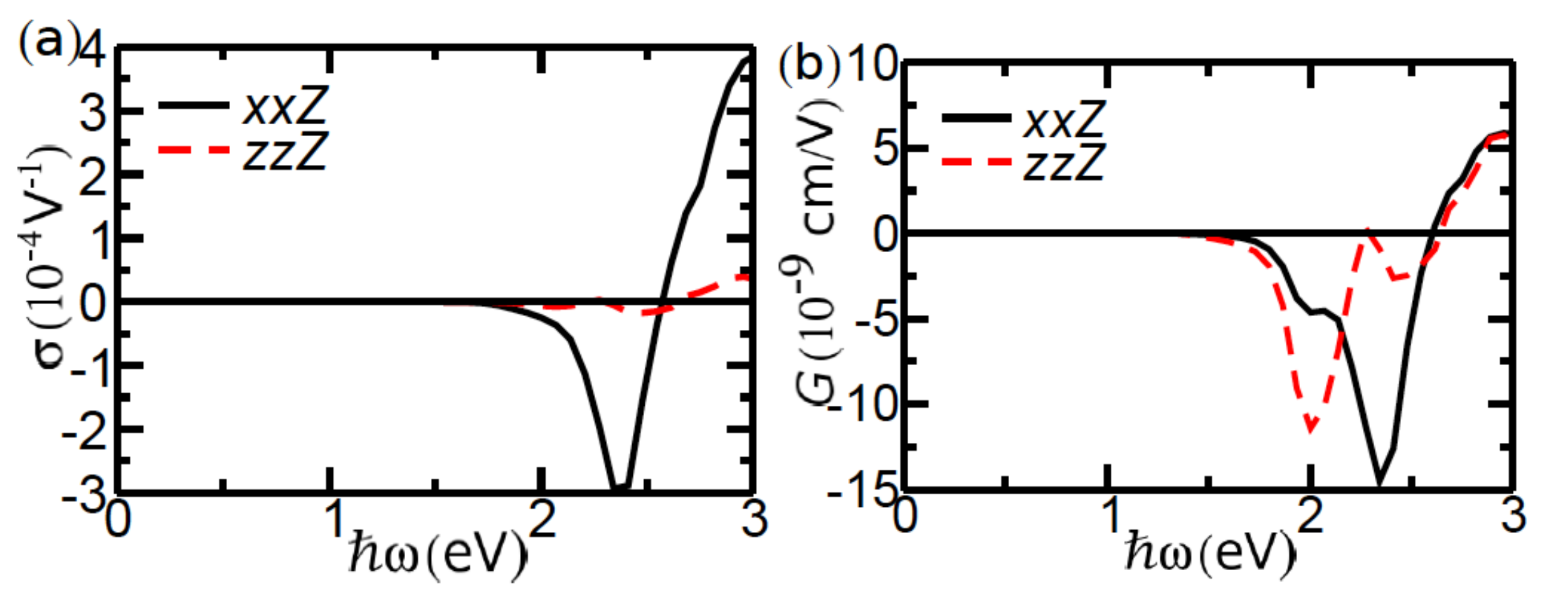}
\caption{(Color online) The (a) shift current susceptibilities and (b) Glass coefficients of the $1\times1\times3$ layered Ni-PTO.
\vspace{-1.5em}
\label{113}}
\end{figure}
%
\begin{figure*}[t]
\centering
\vspace{-1.0em}
\includegraphics[width=\textwidth]{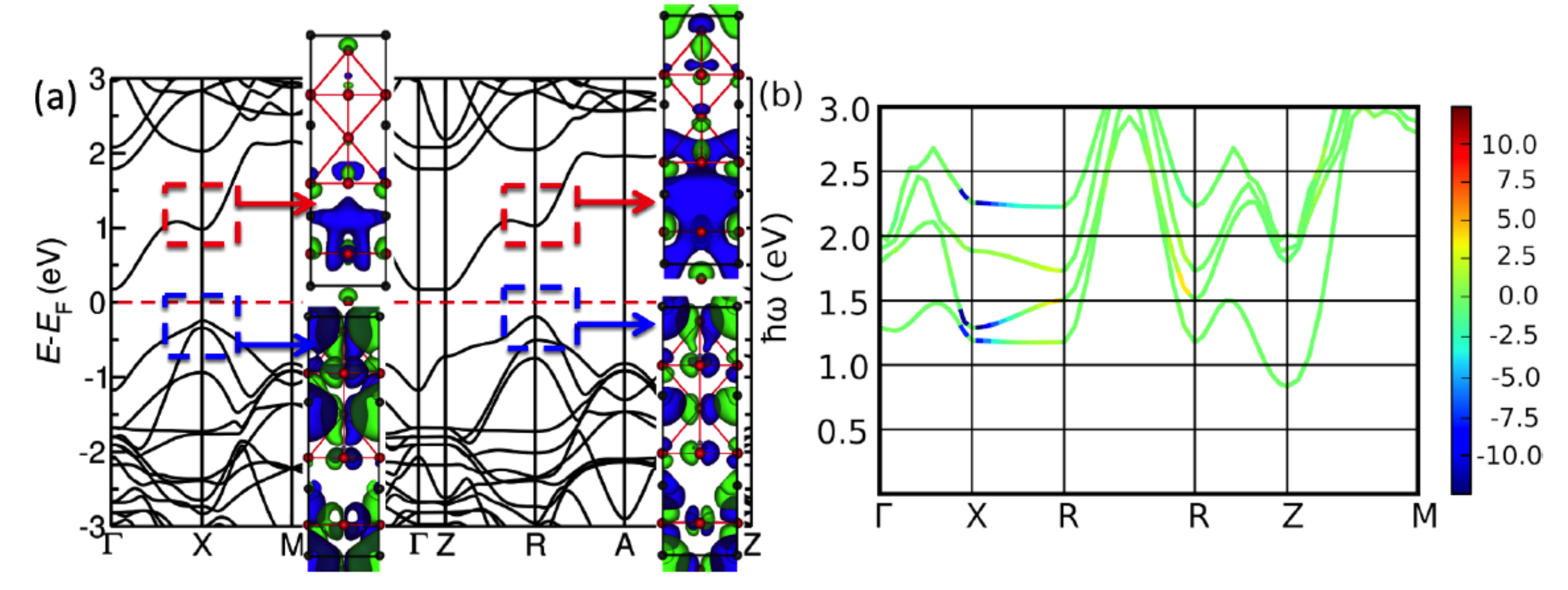}
\caption{(Color online) (a) The band structure and real-space wavefunction isosurfaces corresponding to the states indicated by the dashed rectangular regions of the $1\times1\times3$ layered Ni-PTO. (b) The $k$-resolved shift current for $N$=3. The bands describe the transitions, with the color giving the value of the shift current response (A/V$^{2}$). The near-gap response is dominated by the region around the line from X to R, and changes direction along this path.
\label{band-structure}}
\vspace{-1.5em}
\end{figure*}
A previously developed approach that yields good agreement with experiment for shift current magnitude and spectral profile was used to calculate the shift current.~\cite{Young12p116601, Young12p236601}
The QUANTUM-ESPRESSO code was used to perform density functional theory (DFT) calculations with the local density approximation (LDA) functional~\cite{Giannozzi09p395502, Kohn65pA1133, Perdew81p5048}.
All elements are represented by norm-conserving, optimized nonlocal pseudopotentials~\cite{Rappe90p1227}.
For structural optimizations, at least $6\times6\times6$ Monkhorst-Pack $k$-point grid~\cite{Monkhorst76p5188} was used while for the self-consistent and non-self-consistent calculations, finer $k$-point grids up to $40\times40\times40$ were used to get a well-converged shift current response.
The DFT+$U$ method was used to improve the description of $d$-orbital electrons, with Hubbard $U$ parameterized by the linear-response approach~\cite{Cococcioni05p035105}.
The calculated $U$ values are $\approx$ 4.7 and 8.9 eV for Ti and Ni.
%
%
All photon energies are shifted by 1.05 eV so that PTO is at its experimental band gap (3.6 eV).

%

%
Tetragonal PTO belongs to the $P4mm$ space group, corresponding to the $C_{4v}$ point group, whose third rank shift current response tensor must have the form
\begin{align}
&\sigma=\begin{bmatrix}
                0 &0& 0&0&\sigma_{zyY}&0\\
                0&0&0&\sigma_{zyY}&0&0 \\
                \sigma_{xxZ}&\sigma_{xxZ}&\sigma_{zzZ}&0&0&0
             \end{bmatrix},
\end{align}
%
where the lower and upper case letters represent directions of the light polarization and photocurrent, respectively.  We choose our polarization direction to be normal to the Ni layers, so that the systems analyzed have the same symmetry properties, and (for unpolarized light) only  $\sigma_{xxZ}$ and $\sigma_{zzZ}$ are relevant.  

We first consider a $1\times1\times 3$ supercell.  The band gap is reduced due to the introduction by nickel of a low-lying conduction band (CB) into the bulk PTO gap; as a result, the band-edge electronic transitions, which will be responsible for the low energy shift current response, occur only between the valence bands (VBs) and this particular CB. Fig.~\ref{113} shows the calculated tensor elements of the shift current susceptibility and Glass coefficient, which show response well within the visible range, as expected.   
Additionally, the shift current induced by the perpendicularly polarized light ($\sigma_{xxZ}$) is much stronger than that induced by the parallel polarized light  ($\sigma_{zzZ}$) because of their different absorption efficiencies.
Furthermore, mapping of the $k$-resolved shift current strength (Fig.~\ref{band-structure}) shows that the electronic transitions occurring near X(0, 0.5, 0) and R(0, 0.5, 0.5) $k$-points and along the X-R line induce the most substantial shift current responses and are thus the most important for the band-edge electronic transitions.
However, the contribution from the ends of this region largely counteract each other, as the shift vector changes direction along this line, suggesting that the total response has the potential to be much stronger than what appears in Fig.~\ref{113}.

\begin{figure}[b]
\vspace{-1.0em}
\centering
\includegraphics[width=0.9\textwidth]{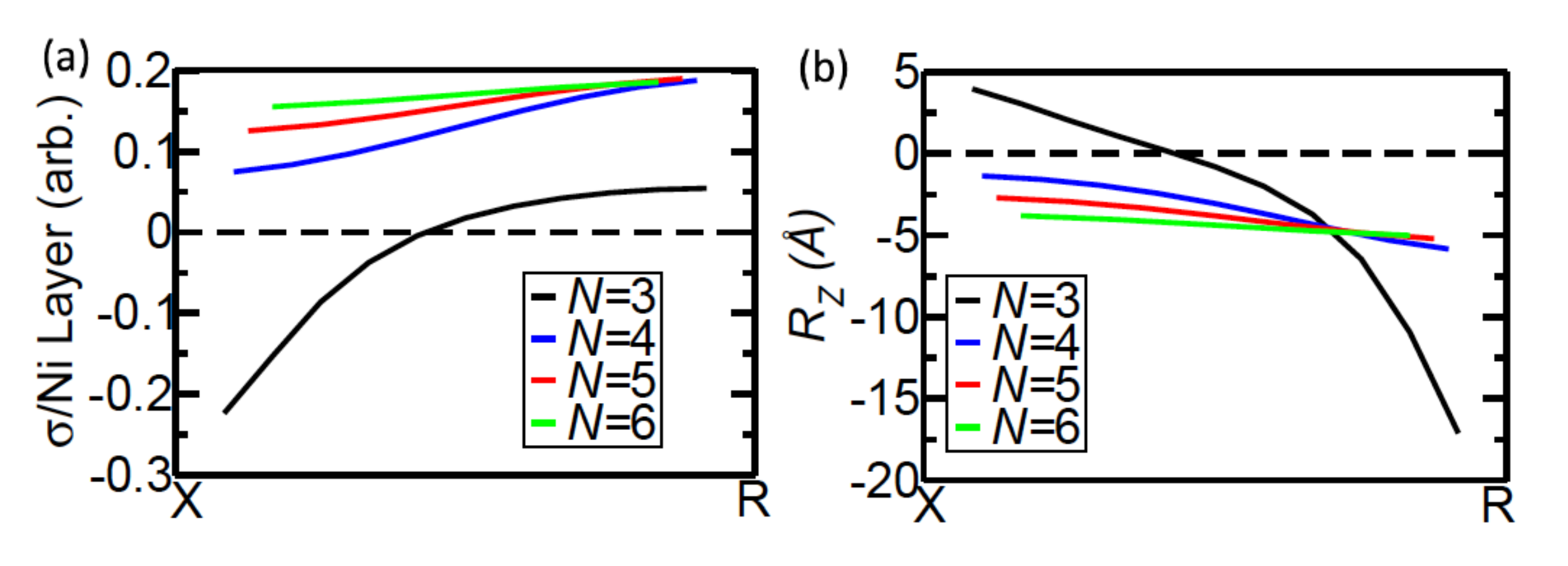}
\caption{(Color online) The shift current susceptibilities ($\sigma_{xxZ}$) and the shift vector for the $k$ points along the X$-$R line in the Brillouin zone of the $1\times1\times N$ layered Ni-PTO.
\label{nl_sc_sv}}
\vspace{-2.0em}
\end{figure}
%
%
\begin{figure*}[t]
\centering
\includegraphics[width=\textwidth]{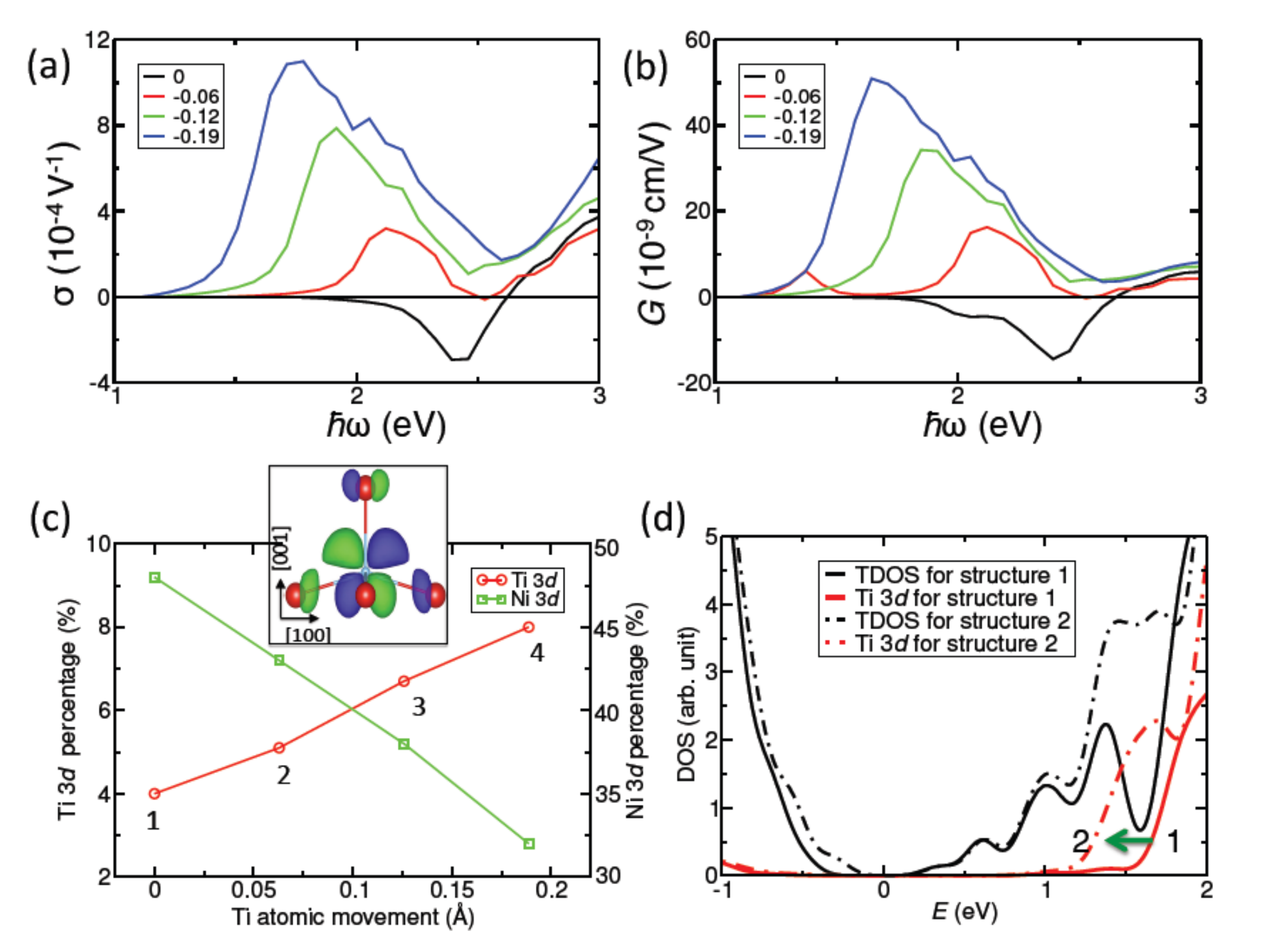}
\caption[.]{(Color online) The (a) shift current susceptibilities ($\sigma_{xxZ}$), (b) Glass coefficients ($G_{xxZ}$), (c) percentages of the Ti and Ni 3$d$ orbitals at the CB ($\sum_{\substack{m}}\big\lvert \langle\psi_{\rm CB,X}\rvert\phi_{32m}\rangle\big\rvert^{2}$) at X $k$ point, and (d) total and projected density of states, of the $1\times1\times3$ layered Ni-PTO with different Ti atomic movements ({\AA}) with respect to their relaxed positions. The inset in (c) shows the orbital character of the Ti-O orbital overlap at the CB. When Ti moves towards the O$_{4}$ plane, the Ti-O interaction becomes more of nonbonding and less of antibonding character.
\label{Ti-move}}
\vspace{-1.5em}
\end{figure*}
To understand this, we consider the details of the electronic states at the points X and R.
Real-space  wavefunction distribution analysis for the highest VB shows that it is essentially delocalized at both X and R, with the wavefunction extending to the whole supercell because of the overlap among the Pb 6$s$, Ti 3$d$, Ni 3$d$, and O 2$p$ orbitals (Fig.~\ref{band-structure}). 
However, the wavefunction distribution for the lowest CB shows substantial difference between X and R.
At X, it arises mainly from the Ni 3$d_{x^{2}-y^{2}}$ and the nearby Pb 6$p$ orbitals, and is more localized, whereas at R it becomes much more delocalized, with a larger orbital contribution from the Pb 6$p$, Ti 3$d_{x^{2}-y^{2}}$ and Ti 3$d_{z^{2}}$ orbitals that are away from Ni.
The local structural environment around Ni is different from those around Pb and Ti, with a square planar symmetry around Ni, but a strong structural asymmetry along the polar axis around Pb and Ti.
The change of the orbital character and the local structural environment can have two effects on the shift vector.
First, compared to the Ni 3$d_{x^{2}-y^{2}}$ orbital, the Pb 6$p$ and Ti 3$d_{x^{2}-y^{2}}$ orbitals are more delocalized, while the Ti 3$d_{z^{2}}$ orbitals extend the wavefunction along the shift current direction; this facilitates the motion of the shift current carriers.
Correspondingly, the shift vector magnitude at R is much larger than that at X (Fig.~\ref{nl_sc_sv}).
Second, it opens up the way for the sign change of the shift vector at R from that at X~\cite{Young12p116601}.
Therefore, if the orbital contribution of the Ti 3$d$ and Pb 6$p$ orbitals to the lowest CB increases, not only is the shift vector magnitude enhanced, but also the orbital character similarity between the X- and R-regions becomes stronger.
Accordingly, the shift current direction at the X-region changes to the same direction as the shift current at the R-region, largely eliminating the cancellation of counter propagating currents.
This means that we should be able to engineer shift current direction and magnitude through the adjustment of the orbital character contribution to the band-edge electronic transitions.

To verify this, we engineer the shift current response by changing the local structure.
Basically, the bottom of the CB is composed of the Ni 3$d$, Ti 3$d$, Pb 6$p$, and O 2$p$ orbitals, with the Ni 3$d$ orbitals making a major contribution (Fig. S1 in the Supplementary).
Since the  peak of the Ti 3$d$ orbitals is located at a higher energy than that of the Ni 3$d$ peak, the contribution of the Ti 3$d$ states to the band-edge electronic transitions can be enhanced by a downshift of the Ti 3$d$ orbitals at the CB.
Therefore, we move Ti atoms antiparallel to the overall polarization, reducing, but not entirely eliminating the local Ti off-center displacements found in the relaxed structure.
Clearly, the shift current direction changes and its magnitude is substantially enhanced by moving the Ti sublattices antiparallel to the overall polarization (Fig.~\ref{Ti-move}).
The shift current onset energy also moves downwards as the Ti sublattice artificial displacement antiparallel to the polarization is increased.
The DOS analysis shows that the contribution of the Ti 3$d$ (and also Pb 6$p$, not shown) orbitals to the lowest CB increases while the proportion of the Ni 3$d$ orbitals decreases steadily with Ti atomic movements (Fig.~\ref{Ti-move}).
When the Ti off-center displacements and the lattice distortions are reduced, the antibonding character of the Ti-O orbital overlap is reduced~\cite{Eng03p94}.
This downshifts the Ti 3$d$ orbitals at the CB and  leads to an enhancement of the Ti 3$d$ (Pb 6$p$) states contributing to the lowest CB.
At a certain point,  the shift vector at the X-region changes sign, and the collective shift current responses for  the X- and R-regions add up constructively.
\begin{figure}[b]
\vspace{-1.0em}
\centering
\includegraphics[width=0.8\textwidth]{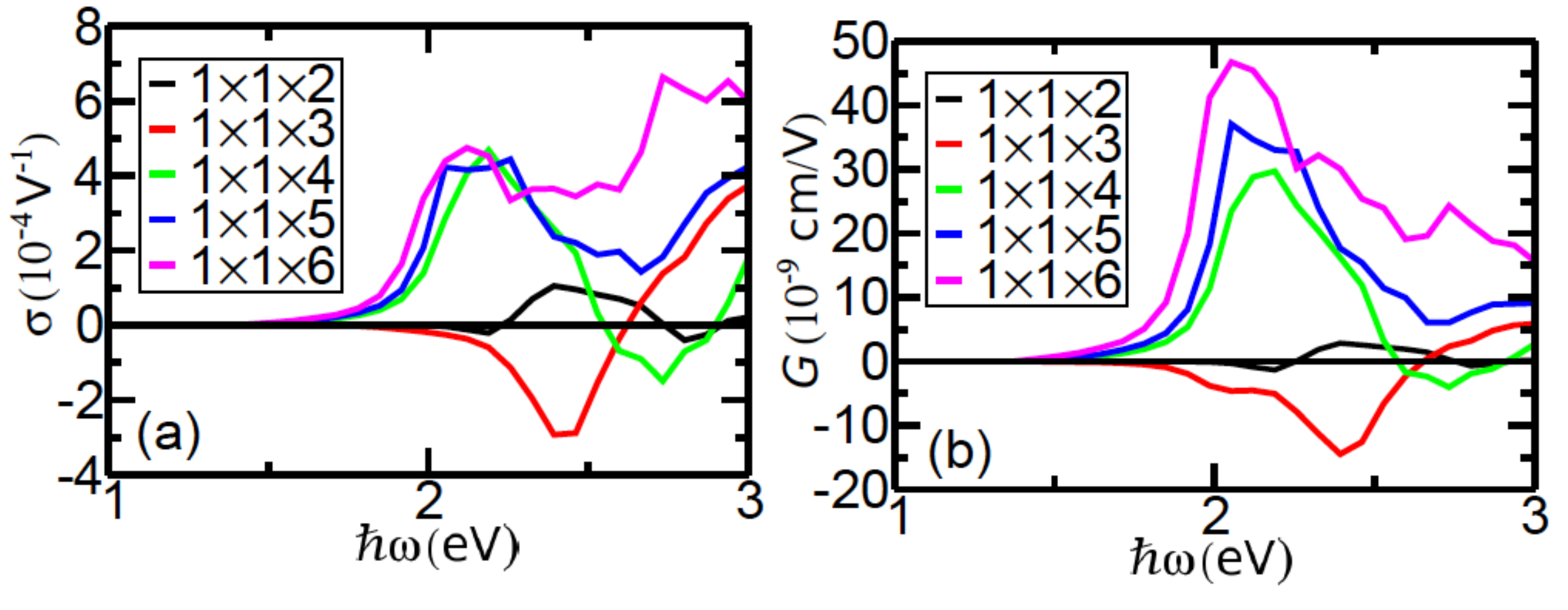}
\caption{(Color online) The (left) shift current susceptibilities ($\sigma_{xxZ}$) and (right) Glass coefficients ($G_{xxZ}$) of the layered Ni-PTO with varying superlattice periodicity. 
\label{11n}}
\end{figure}

\begin{figure}[t]
\centering
\includegraphics[width=0.8\textwidth]{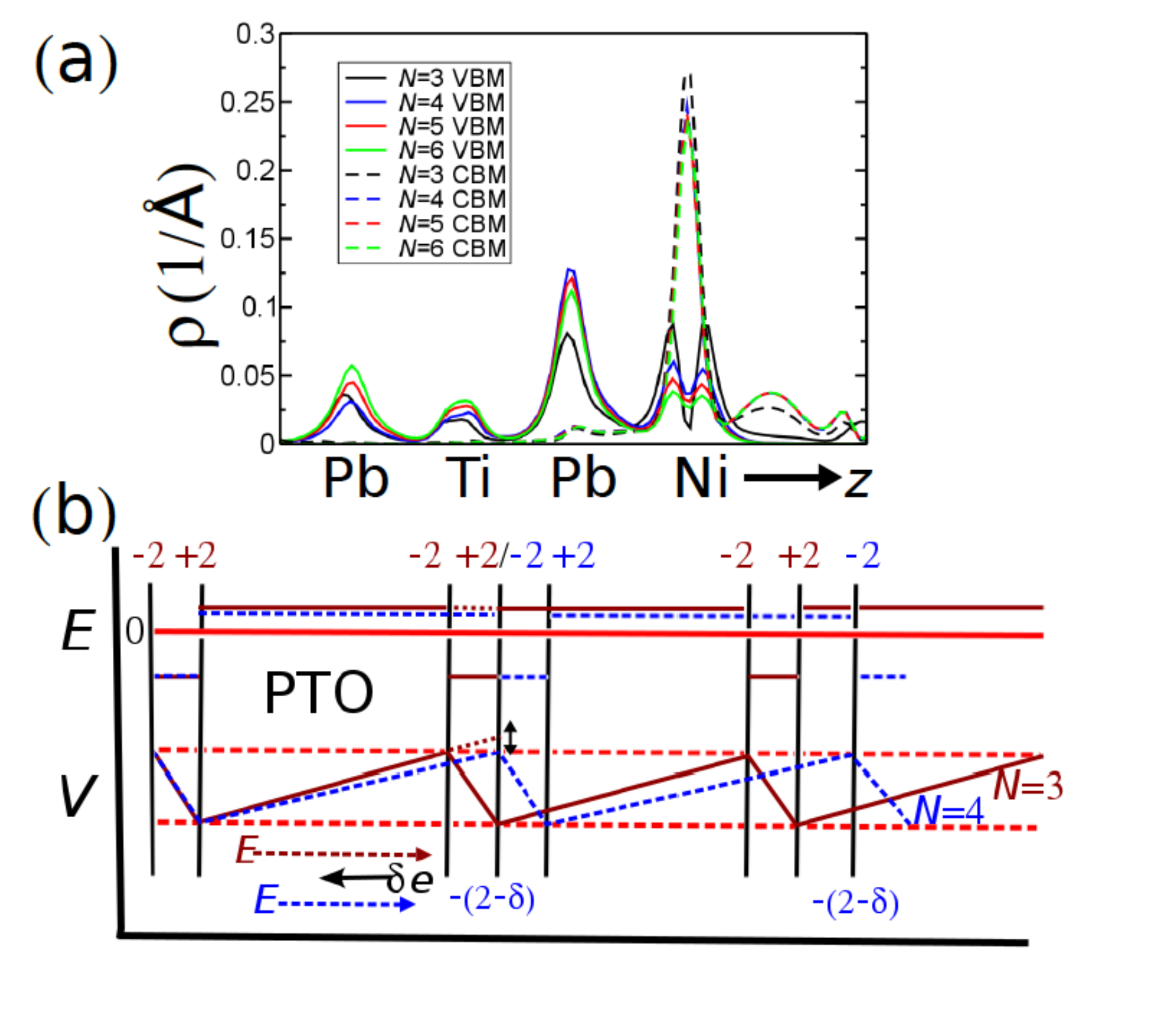}
\caption[.]{(Color online) (a) The real-space probability density distribution summed over the Cartesian $x$ and $y$ coordinates ($\rho(z)$=$\int\!{\rm d}x{\rm d}y \big\lvert\langle\vv{r}\rvert\psi_{n_{0},\rm R}\rangle\big\rvert^{2}$) for the VBM and CBM states at R $k$ point. $N$ is the number of layers. (b)The electric field and potential inside Ni-PTO for $N$=3 and 4. The Ni-$V_{\rm O}$ substitution results in adjacent planes of charge. As layers increase, stronger screening over a greater distance is generated in response to the potential changes introduced by this charge separation. The electric field magnitude over the PTO layers is approximately $\frac{2e}{(2N-1)\epsilon_{0}}$, decreasing as $N$ increases.
\label{probability-electrostatic}}
\vspace{-1.5em}
\end{figure}

It is clear that structural distortion of the bulk PTO can significantly affect the contribution of states dominated by the nickel layer. Such manipulation of the atomic configuration is artificial, but suggests that altering the number of bulk PTO layers may have a significant effect.  This is confirmed by the response for thicker Ni-PTO supercells with one NiO$_{2}$ plane per supercell (Fig.~\ref{11n}); contrary to naive expectations, reducing the Ni fraction actually increases the effectiveness of substitution, increasing the response and reducing the band gap as the number of layers is increased.
The proximate reason for the change in response becomes clear upon looking at the response along the X-R line (Fig.~\ref{nl_sc_sv}). The dramatic change going from $1\times1\times3$ to $1\times1\times4$ is due to the disappearance of the sign change versus $k$ point in the shift current response response, and as the number of PTO layers is further increased, the magnitude of the shift current further increases.
To investigate further, we plot the layer-averaged probability density as a function of layer normal coordinate for the relevant VBM and CBM states (Fig.~\ref{probability-electrostatic}). Going from  $N$=3 to $N$=4, these states, especially the VB state, becomes more delocalized, corresponding to the enhancement of the shift current magnitude. Additionally, this delocalization is only moderately enhanced from $N$=4 to $N$=6, resulting in the relatively flat progression of shift vector observed for $N>3$. 

This state delocalization can ultimately be understood from the conceptual picture of charge separation introduced by the Ni-$V_{\rm O}$ substitution. In Ni-PTO, the Ni$_{\rm Ti}^{''}$ layer has $-2$ charge, adjacent to the $V_{\rm O}^{\dotr\dotr}$ layer with $+2$ charge. Approximating these as simple charged planes, this results in a periodic series of regions with electric field, similar to those responsible for the polar catastrophe in LaAlO$_3$/SrTiO$_{3}$~\cite{Ohtomo04p423}. However, in this case, the distance between each dipolar pair of planes and the next is not constant. As $N$ increases, the electric field in the PTO layers becomes smaller under short-circuit boundary conditions, and bound charges face a stronger impetus to screen the field over a greater length, delocalizing electron densities (Fig.~$\ref{probability-electrostatic}$).

The foregoing holds important lessons for engineering BPV materials.  For one, it indicates that the problem of band gap and shift current generating states can be partially separated. Nickel substitution provides lower-energy CB states, that, while tending to localization along the layer normal direction, which is not favorable for shift current, nonetheless may participate in transitions with what are essentially PTO states that are delocalized and thus shift current favorable.  We have thus chosen a material with robust shift current potential but too large a gap (PTO), and introduced a modification that solves the latter without losing the former. Second, it emphasizes the importance of not only the magnitude of shift vector as a factor in response, but consistency in direction,  which is sensitive to relatively small structural changes, and therefore amenable to manipulation. Third, 
the electrostatic effects of the Ni-$V_{\rm O}$ substitution on the electronic structure of bulk PTO are pronounced and most apparent for lower nickel fractions.  

Finally, we wish to point out an additional potential benefit of the above structures.  Recently, the importance of conventional transport characteristics has been highlighted~\cite{Bhatnagar13p1}.  Essentially, the performance of BPV materials depends not only on their current generation capability, but on the photovoltage they can sustain.  This latter quantity is determined by resistance of current leaking back through the material, and the design of useful devices will require control over conventional conductivity.  In BiFeO$_{3}$, domain walls can serve as barricades against such leakage, substantially increasing the resistance and, consequently, maximum photovoltage.  We note that the present system similarly features a nanoscale heterogeneous layered structure, and, in particular, that an additional consequence of increasing $N$ is delocalizing electronic states.  In this sense, the structure may be viewed as a nanoscale composite with alternating photocurrent generating layers and insulating layers, ideal for a BPVE device.

\begin{acknowledgments}
F. W. was supported by the National Science Foundation, under Grant DMR-1124696. S. M. Y. was supported by the Department of Energy Office of Basic Energy Sciences, under Grant DE-FG02-07ER46431. F. Z. was supported by the National Science Foundation, under Grant CMMI-1334241. I. G. was supported by the office of Naval Research, under Grant N00014-12-1-1033. A. M. R. was supported by the Office of Naval Research under Grant No. N00014-11-1-0664. Computational support was provided by the High-Performance Computing Modernization Office of the Department of Defense and the National Energy Research Scientific Computing Center of the Department of Energy.
\end{acknowledgments}


\begin{thebibliography}{21}
\bibitem{Maeda06p295}K. Maeda, K. Teramura, D. Lu, T. Takata, N. Saito, Y. Inoue, and K. Domen, Nature $\bf{440}$, 295 (2006).
\bibitem{Choi09p63}T. Choi, S. Lee, Y. Choi, V. Kiryukhin, and S.-W. Cheong, Science $\bf{324}$, 63 (2009).
\bibitem{Kudo09p253}A. Kudo and Y. Miseki, Chem. Soc. Rev. $\bf{38}$, 253 (2009).
\bibitem{Wang12p476}F. Wang, C. Di Valentin, and G. Pacchioni, ChemCatChem $\bf{4}$, 476 (2012).
\bibitem{Glass74p233}A. M. Glass, D. von der Linde, and T. J. Negran, Appl. Phys. Lett. $\bf{25}$, 233 (1974).
\bibitem{Kraut79p1548}W. Kraut and R. von Baltz, Phys. Rev. B $\bf{19}$, 1548 (1979).
\bibitem{Chynoweth56p705}A. G. Chynoweth, Phys. Rev. $\bf{102}$, 705 (1956).
\bibitem{Shockley61p510}W. Shockley and H. Queisser, J. Appl. Phys. $\bf{32}$, 510 (1961).
\bibitem{Ji10p1763}W. Ji, K. Yao, and Y. C. Liang, Adv. Mater. $\bf{22}$, 1763 (2010).
\bibitem{Baltz81p5590}R. von Baltz and W. Kraut, Phys. Rev. B $\bf{23}$, 5590 (1981).
\bibitem{Young12p116601}S. M. Young and A. M. Rappe, Phys. Rev. Lett. $\bf{109}$, 116601 (2012).
\bibitem{Bennett08p17409}J. W. Bennett, I. Grinberg, and A. M. Rappe, J. Am. Chem. Soc. $\bf{130}$, 17409 (2008).
\bibitem{Gou11p205115}G. Y. Gou, J. W. Bennett, H. Takenaka, and A. M. Rappe, Phys. Rev. B $\bf{83}$, 205115 (2011).
\bibitem{Young12p236601}S. M. Young, F. Zheng, and A. M. Rappe, Phys. Rev. Lett. $\bf{109}$, 236601 (2012).
\bibitem{Giannozzi09p395502}P. Giannozzi, S. Baroni, N. Bonini, M. Calandra, R. Car, C. Cavazzoni, D. Ceresoli, G. L. Chiarotti, M. Cococcioni, I. Dabo, A. D. Corso, S. de Gironcoli, S. Fabris, G. Fratesi, R. Gebauer, U. Gerstmann, C. Gougoussis, A. Kokalj, M. Lazzeri, L. Martin-Samos, N. Marzari, F. Mauri, R. Mazzarello, S. Paolini, A. Pasquarello, L. Paulatto, C. Sbraccia, S. Scandolo, G. Sclauzero, A. P. Seitsonen, A. Smogunov, P. Umari, and R. M. Wentzcovitch, J. Phys.: Condens. Matter $\bf{21}$, 395502 (2009).
\bibitem{Kohn65pA1133}W. Kohn and L. J. Sham, Phys. Rev. $\bf{140}$, A1133 (1965).
\bibitem{Perdew81p5048}J. P. Perdew and A. Zunger, Phys. Rev. B $\bf{23}$, 5048 (1981).
\bibitem{Rappe90p1227}A. M. Rappe, K. M. Rabe, E. Kaxiras, and J. D. Joannopoulos, Phys. Rev. B Rapid Comm. $\bf{41}$, 1227 (1990).
\bibitem{Monkhorst76p5188}H. J. Monkhorst and J. D. Pack, Phys. Rev. B $\bf{13}$, 5188 (1976).
\bibitem{Cococcioni05p035105}M. Cococcioni and S. de Gironcoli, Phys. Rev. B $\bf{71}$, 035105 (2005).
\bibitem{Eng03p94} H. W. Eng, P. W. Barnes, B. M. Auer, and P. M. Woodward, J. Solid State Chem $\bf{175}$, 94 (2003).
\bibitem{Ohtomo04p423}A. Ohtomo, H. Y. Hwang, Nature $\bf{427}$, 423 (2004).
\bibitem{Bhatnagar13p1}A. Bhatnagar, A. R. Chaudhuri, Y. H. Kim, D. Hesse, and M. Alexe, Nat. Commun. $\bf{4}$, 2835 (2013).
\end{thebibliography}
\end{document}